\documentclass[12pt]{article}
\usepackage{graphicx}

\newcommand\pubnumber{}
\newcommand\pubdate{\today}

\def\Title#1{\begin{center} {\Large #1 } \end{center}}
\def\Author#1{\begin{center}{ \sc #1} \end{center}}
\def\Address#1{\begin{center}{ \it #1} \end{center}}

\newcommand\pubblock{\rightline{\begin{tabular}{l} \pubnumber\\
         \pubdate  \end{tabular}}}
\newenvironment{Abstract}{\begin{quotation}  }{\end{quotation}}
\newenvironment{Presented}{\begin{quotation} \begin{center} 
             PRESENTED AT\end{center}\bigskip 
      \begin{center}\begin{large}}{\end{large}\end{center} \end{quotation}}
\def\Acknowledgements{\bigskip  \bigskip \begin{center} \begin{large}
             \bf ACKNOWLEDGEMENTS \end{large}\end{center}}
\def\beq{\begin{equation}}
\def\eeq#1{\label{#1}\end{equation}}
\def\eeqn{\end{equation}}

\def\beqa{\begin{eqnarray}}
\def\eeqa#1{\label{#1}\end{eqnarray}}
\def\eeqan{\end{eqnarray}}

\let\bar=\overbar

\def\D{{\cal D}}

\def\Dslash{\not{\hbox{\kern-4pt $D$}}}
\def\dslash{\not{\hbox{\kern-2pt $\del$}}}

\def\msb{{\bar{\ssstyle M \kern -1pt S}}}

\usepackage{mciteplus}
\usepackage{cite}

\usepackage{xspace}

\usepackage{amsmath} % Adds a large collection of math symbols    
\usepackage{amssymb}                                                                                                                      
\usepackage{lineno}
\usepackage{amsfonts}
\usepackage{upgreek} % Adds in support for greek letters in roman typeset                                                                                                                 

\usepackage{cancel}

\usepackage{ifthen} % for conditional statements
\newboolean{articletitles}
\setboolean{articletitles}{true} % False removes titles in references
\newboolean{uprightparticles}
\setboolean{uprightparticles}{false} %Set to true to get roman particle symbols

\def\lhcb {\mbox{LHCb}\xspace}
\def\ux85 {\mbox{UX85}\xspace}

\def\babar  {\mbox{BaBar}\xspace}

\ifthenelse{\boolean{uprightparticles}}%
{

 \def\Pmu         {\ensuremath{\upmu}\xspace}

 \def\Ppi         {\ensuremath{\uppi}\xspace}

 \def\PDelta      {\ensuremath{\Delta}\xspace}                 
 \def\PXi      {\ensuremath{\Xi}\xspace}                 
 \def\PLambda      {\ensuremath{\Lambda}\xspace}                 
 \def\PSigma      {\ensuremath{\Sigma}\xspace}                 
 \def\POmega      {\ensuremath{\Omega}\xspace}                 
 \def\PUpsilon      {\ensuremath{\Upsilon}\xspace}                 
 
 \def\PB      {\ensuremath{\mathrm{B}}\xspace}                 
                  
 \def\PD      {\ensuremath{\mathrm{D}}\xspace}

 \def\PK      {\ensuremath{\mathrm{K}}\xspace}

 \def\Pe      {\ensuremath{\mathrm{e}}\xspace}

 \def\Pi      {\ensuremath{\mathrm{i}}\xspace}

 \def\Ps      {\ensuremath{\mathrm{s}}\xspace}

}
{

 \def\Pmu         {\ensuremath{\mu}\xspace}

 \def\Ppi         {\ensuremath{\pi}\xspace}

 \mathchardef\PDelta="7101
 \mathchardef\PXi="7104
 \mathchardef\PLambda="7103
 \mathchardef\PSigma="7106
 \mathchardef\POmega="710A
 \mathchardef\PUpsilon="7107
                  
 \def\PB      {\ensuremath{B}\xspace}                 
                  
 \def\PD      {\ensuremath{D}\xspace}

 \def\PK      {\ensuremath{K}\xspace}

 \def\Pe      {\ensuremath{e}\xspace}

 \def\Pi      {\ensuremath{i}\xspace}

 \def\Ps      {\ensuremath{s}\xspace}

}

\def\en         {\ensuremath{\Pe^-}\xspace}   % electron negative (\em is taken)
\def\ep         {\ensuremath{\Pe^+}\xspace}

\def\mup        {\ensuremath{\Pmu^+}\xspace}
\def\mun        {\ensuremath{\Pmu^-}\xspace} % muon negative (\mum is taken)
\def\mumu       {\ensuremath{\Pmu^+\Pmu^-}\xspace}

\def\ellell     {\ensuremath{\ell^+ \ell^-}\xspace}

\def\squark    {\ensuremath{\Ps}\xspace}

\def\pion  {\ensuremath{\Ppi}\xspace}

\def\pip   {\ensuremath{\pion^+}\xspace}
\def\pim   {\ensuremath{\pion^-}\xspace}

\def\kaon  {\ensuremath{\PK}\xspace}
  \def\Kbar  {\kern 0.2em\overline{\kern -0.2em \PK}{}\xspace}

\def\Kz    {\ensuremath{\kaon^0}\xspace}
\def\Kzb   {\ensuremath{\Kbar^0}\xspace}
\def\KzKzb {\ensuremath{\Kz \kern -0.16em \Kzb}\xspace}
\def\Kp    {\ensuremath{\kaon^+}\xspace}
\def\Km    {\ensuremath{\kaon^-}\xspace}

\def\KpKm  {\ensuremath{\Kp \kern -0.16em \Km}\xspace}
\def\KS    {\ensuremath{\kaon^0_{\rm\scriptscriptstyle S}}\xspace} 
\def\KL    {\ensuremath{\kaon^0_{\rm\scriptscriptstyle L}}\xspace} 
\def\Kstarz  {\ensuremath{\kaon^{*0}}\xspace}

\def\Kstar   {\ensuremath{\kaon^*}\xspace}

  \def\Dbar    {\kern 0.2em\overline{\kern -0.2em \PD}{}\xspace}
\def\D       {\ensuremath{\PD}\xspace}

\def\Dz      {\ensuremath{\D^0}\xspace}
\def\Dzb     {\ensuremath{\Dbar^0}\xspace}
\def\DzDzb   {\ensuremath{\Dz {\kern -0.16em \Dzb}}\xspace}
\def\Dp      {\ensuremath{\D^+}\xspace}
\def\Dm      {\ensuremath{\D^-}\xspace}

\def\DpDm    {\ensuremath{\Dp {\kern -0.16em \Dm}}\xspace}

\def\B       {\ensuremath{\PB}\xspace}
  \def\Bbar    {\kern 0.18em\overline{\kern -0.18em \PB}{}\xspace}

\def\Bz      {\ensuremath{\B^0}\xspace}
\def\Bzb     {\ensuremath{\Bbar^0}\xspace}
\def\Bu      {\ensuremath{\B^+}\xspace}

\def\Bp      {\ensuremath{\Bu}\xspace}

\def\Bs      {\ensuremath{\B^0_\squark}\xspace}

  \def\Y#1S{\ensuremath{\PUpsilon{(#1S)}}\xspace}% no space before {...}!

\def\Lbar {\ensuremath{\kern 0.1em\overline{\kern -0.1em\Lambda\kern -0.05em}\kern 0.05em{}}\xspace}

\def\BF         {{\ensuremath{\cal B}\xspace}}

\newcommand{\decay}[2]{\ensuremath{#1\!\to #2}\xspace}         % {\Pa}{\Pb \Pc}

\def\to                 {\ensuremath{\rightarrow}\xspace}

\def\qsq       {\ensuremath{q^2}\xspace}

\def\CP                {\ensuremath{C\!P}\xspace}

\def\AT#1     {\ensuremath{A_{\mathrm{T}}^{#1}}\xspace}           % 2

\def\C#1      {\ensuremath{\mathcal{C}_{#1}}\xspace}                       % 9
\def\Cp#1     {\ensuremath{\mathcal{C}_{#1}^{'}}\xspace}                    % 7
\def\Ceff#1   {\ensuremath{\mathcal{C}_{#1}^{\mathrm{(eff)}}}\xspace}        % 9  
\def\Cpeff#1  {\ensuremath{\mathcal{C}_{#1}^{'\mathrm{(eff)}}}\xspace}       % 7
\def\Ope#1    {\ensuremath{\mathcal{O}_{#1}}\xspace}                       % 2
\def\Opep#1   {\ensuremath{\mathcal{O}_{#1}^{'}}\xspace}                    % 7

\newcommand{\tev}{\ensuremath{\mathrm{\,Te\kern -0.1em V}}\xspace}
\newcommand{\gev}{\ensuremath{\mathrm{\,Ge\kern -0.1em V}}\xspace}
\newcommand{\mev}{\ensuremath{\mathrm{\,Me\kern -0.1em V}}\xspace}
\newcommand{\kev}{\ensuremath{\mathrm{\,ke\kern -0.1em V}}\xspace}
\newcommand{\ev}{\ensuremath{\mathrm{\,e\kern -0.1em V}}\xspace}
\newcommand{\gevc}{\ensuremath{{\mathrm{\,Ge\kern -0.1em V\!/}c}}\xspace}
\newcommand{\mevc}{\ensuremath{{\mathrm{\,Me\kern -0.1em V\!/}c}}\xspace}
\newcommand{\gevcc}{\ensuremath{{\mathrm{\,Ge\kern -0.1em V\!/}c^2}}\xspace}
\newcommand{\gevgevcccc}{\ensuremath{{\mathrm{\,Ge\kern -0.1em V^2\!/}c^4}}\xspace}
\newcommand{\mevcc}{\ensuremath{{\mathrm{\,Me\kern -0.1em V\!/}c^2}}\xspace}

\def\invfb   {\ensuremath{\mbox{\,fb}^{-1}}\xspace}

\def\deriv {\ensuremath{\mathrm{d}}}

\def\gsim{{~\raise.15em\hbox{$>$}\kern-.85em
          \lower.35em\hbox{$\sim$}~}\xspace}
\def\lsim{{~\raise.15em\hbox{$<$}\kern-.85em
          \lower.35em\hbox{$\sim$}~}\xspace}

\newcommand{\Imag}{\ensuremath{\mathcal{I}m}\xspace}

\def\tell1  {TELL1\xspace}
\def\ukl1   {UKL1\xspace}

\usepackage{hyperref}    % Hyperlinks in references   

\begin{document}
\begin{titlepage}
\pubblock

\vfill
\Title{Summary of the CKM 2014 working group on rare decays}
\vfill
\Author{Thomas~Blake$^{1}$, Akimasa~Ishikawa$^{2}$, David~M.~Straub$^{3}$}
\Address{$^1$\,Department of Physics, Univeristy of Warwick, UK \\  $^2$\,Department of Physics, Tohoku University, Japan \\ $^3$\,Universe Cluster, TU Munich, Germany}
\vfill
\begin{Abstract}
\noindent Rare flavour changing neutral current decays of strange, charm and beauty hadrons have been instrumental in building up a picture of flavour in the Standard Model. Increasingly precise measurements of these decays allow to search for deviations from predictions of the Standard Model that would be associated to contributions from new particles that might arise in extensions of the Standard Model.  In this summary, an overview of recent experimental results and theoretical predictions is given. The new physics sensitivity and prospects for the different observables is also addressed. 
\end{Abstract}
\vfill
\begin{Presented}
 8th International Workshop on the CKM Unitarity Triangle (CKM 2014), Vienna, Austria, September 8-12, 2014
\end{Presented}
\vfill
\end{titlepage}
\def\thefootnote{\fnsymbol{footnote}}
\setcounter{footnote}{0}

\section{Introduction}

Weak flavour-changing neutral current (FCNC) decays of hadrons are loop and CKM suppressed in the Standard Model (SM). These \textit{rare decays} played an important role in formulating the Cabibbo-Kobayashi-Maskawa picture of quark mixing. They continue to play an important role in theory and experiment as they are potentially sensitive to contributions from heavy physics beyond the SM. Compared to meson-antimeson mixing, rare decays provide a much larger number of observables susceptible to such new physics effects. The flavour, \CP, and chirality structure of new interactions can be probed by branching fractions and various decay asymmetries in decays of strange, charm, and beauty hadrons. In the SM, lepton flavour is conserved by accidental symmetries and searches for charged lepton flavour violation can also provide important null tests of the SM.

The past two years have seen a wealth of new measurements of rare $b$- and $c$-hadron decay processes by the ATLAS, CMS and LHCb experiments at the LHC and from the legacy datasets of the \babar and Belle experiments.  There have also been updated limits on lepton flavour violating processes, including results from the MEG experiment and first results from a pilot run of the KOTO experiment. An overview of the experimental and theoretical status of these measurements is given below. 

\section{Rare $b$-hadron decays}

Rare FCNC decays of $b$-hadrons are described by the weak effective Hamiltonian
\begin{equation}
\label{eq:Heff}
{\cal H}_\text{eff} = - \frac{4\,G_F}{\sqrt{2}} V_{tb}V_{tq}^* \frac{e^2}{16\pi^2}
\sum_i
(C_i Q_i + C'_i Q'_i) + \text{h.c.} ,
\end{equation}
where $q=d,s$ for processes based on the quark-level $b\to d,s$ transition.
Among the dimension six operators contributing to these transitions, the operators most sensitive to new physics effects are
\begin{align}
Q_7^{q(\prime)} &= \frac{m_b}{e}
(\bar{q} \sigma_{\mu \nu} P_{R(L)} b) F^{\mu \nu},
\label{eq:O7}
&
Q_8^{q(\prime)} &= \frac{m_b}{e}
(\bar{q} \sigma_{\mu \nu} P_{R(L)} T^a b) G^{a\mu \nu},
\\
Q_9^{q(\prime)} &= 
(\bar{q} \gamma_{\mu} P_{L(R)} b)(\bar{\ell} \gamma^\mu \ell)\,,
&
Q_{10}^{q(\prime)} &=
(\bar{q} \gamma_{\mu} P_{L(RR} b)( \bar{\ell} \gamma^\mu \gamma_5 \ell)\,,\label{eq:O10}
\\
Q_{S}^{q(\prime)} &=
(\bar{q}  P_{R(L)} b)( \bar{\ell} \ell)\,,
&
Q_{P}^{q(\prime)} &=
(\bar{q}  P_{R(L)} b)( \bar{\ell} \gamma_5\ell)\,,
&\\
Q_{L(R)}^q &=
(\bar{q} \gamma_{\mu} P_{L(R)} b)( \bar{\nu} \gamma^\mu P_L \nu)\,.
\end{align}
Here, $P_{L, R}$ denote left and right-hand chirality projections and $F^{\mu \nu}$ and $G^{a \mu \nu}$ are the electromagnetic and chromagnetic field strength tensors, respectively. 
The electromagnetic and chromomagnetic dipole operators ($Q_7^{(\prime)}$ and $Q_8^{(\prime)}$) contribute to radiative and semileptonic decays. The semileptonic operators $Q_{10,S,P}^{(\prime)}$ contribute to leptonic and semileptonic decays, the operators $Q_{9}^{(\prime)}$ only to semileptonic decays, and the operators $Q_{L,R}$ to decays with neutrinos in the final state.

\subsection{Leptonic decays} 

 In addition to the loop and CKM suppression,
the rare decays \decay{\Bs}{\mumu} and \decay{\Bz}{\mumu} 
% are loop (GIM), helicity, and CKM
are strongly helicity suppressed in the SM. Recently, calculations of NLO electroweak corrections \cite{Bobeth:2013tba} and NNLO QCD corrections \cite{Hermann:2013kca} have reduced the uncertainty of the Standard Model predictions, resulting in time integrated branching fractions of~\cite{Bobeth:2013uxa} 
\begin{align}
\BF(\decay{\Bs}{\mumu})_{\rm SM} & = (3.65\pm 0.23)\times 10^{-9} ~~{\rm and} \\
\BF(\decay{\Bz}{\mumu})_{\rm SM} & = (1.06\pm 0.09)\times 10^{-10}  ~.
\end{align} 
The uncertainties on the SM values are now dominated by our knowledge of the CKM matrix elements and the \Bz and \Bs decay constants ($f_{\Bz}$ and $f_{\Bs}$), determined by Lattice QCD calculations~\cite{Aoki:2013ldr}.

Prior to data taking at the LHC, no evidence had been seen for either decay. Using their full datasets from 2011 and 2012, the CMS and LHCb experiments have reported evidence for the \decay{\Bs}{\mumu} at the level of four standard deviations ($\sigma$)~\cite{Chatrchyan:2013bka,LHCb-PAPER-2013-046}.  A combined analysis of the two datasets, shown for the first time at CKM 2014~\cite{CMS:2014xfa}, results in 
 \begin{align}
\BF(\decay{\Bs}{\mumu}) & =   (2.8^{\,+0.7}_{\,-0.6}) \times 10^{-9} ~~{\rm and} \\
\BF(\decay{\Bz}{\mumu}) & =   (3.9^{\,+1.6}_{\,-1.4}) \times 10^{-10} ~. \\
\end{align} 
The decay \decay{\Bs}{\mumu} is observed at more than $5\sigma$ and evidence for the \Bz decay is seen at more than $3\sigma$. The branching fraction of the \Bz mode is consistent with the SM, albeit only at $2.3\sigma$.

The observation of $\decay{\Bs}{\mumu}$,  at a rate consistent with the SM expectation, sets strong constraints on new physics models predicting a large violation of the helicity suppression, e.g.\ the MSSM with large $\tan\beta$ (see e.g.~\cite{Altmannshofer:2012ks}). Future precision measurements will be important to probe models where the helicity suppression is active. In this case, the leptonic decays are complementary to -- and in general theoretically cleaner than -- semileptonic decays.
The ratio of the $\decay{\Bs}{\mumu}$ to $\decay{\Bz}{\mumu}$ branching fractions is an important test of the hypothesis of Minimal Flavour Violation (MFV) \cite{Buras:2003td}.

\subsection{Semileptonic decays}\label{sec:sl}

\subsubsection*{Exclusive semileptonic decays}

Predictions of observables in exclusive rare semi-leptonic $B$ decays face two main challenges. First, determining the QCD form factors of the heavy-to-light meson transition. Second, computing additional hadronic contributions that arise from a hadronic interaction mediated by a virtual photon. The theoretical methods to tackle these challenges are very different for the low and the high end of dilepton invariant mass squared, $q^2$. At intermediate $q^2$, the theoretical situation is even less clear, narrow charmonium resonances lead to a breakdown of quark-hadron duality.

The form factors for the $B\to K$, $B\to K^*$, and $B_s\to \phi$ transitions have recently been computed at high $q^2$ using lattice QCD \cite{Bouchard:2013eph,Horgan:2013hoa}. At low $q^2$, the most accurate predictions are obtained from light-cone sum rules \cite{Ball:2004ye,Ball:2004rg,Khodjamirian:2010vf}. In the case of $B\to K^*\mu^+\mu^-$ decays, instead of using the full QCD form factors, an alternative approach is to use the soft form factors in the heavy quark limit and to construct observables that are independent of these soft form factors at leading order \cite{Descotes-Genon:2013vna,Jager:2012uw,Descotes-Genon:2014uoa}.

Whilst branching fraction predictions are  plagued by sizeable hadronic uncertainties, precise predictions are possible for ratios of decay rates, such as isospin asymmetries, direct \CP asymmetries or ratios involving decays that differ by the flavour of the lepton in the final state. For example, after correcting for final state radiation and bremstrahlung from the electrons, the ratio 

\begin{equation} 
R_{\rm K} = {\Gamma[\decay{\Bp}{\Kp\mup\mun}]}/{\Gamma[\decay{\Bp}{\Kp\ep\,\en\,}]}
\label{eq:RK}
\end{equation} 

\noindent  in the range $1 < \qsq < 6\gev^{2}/c^{4}$ is expected to be unity up-to $10^{-3}$ in the SM~\cite{Bobeth:2007dw}. Using the dataset collected in 2011 and 2012, the \lhcb experiment measures~\cite{LHCb-PAPER-2014-024}

\begin{displaymath} 
R_{\rm K} = 0.745 {}^{+0.090}_{-0.074} {}^{+0.035}_{-0.035}~.
\end{displaymath} 

\noindent This measurement is discrepant from the SM expectation of unity by $2.6\sigma$. 

Since a violation of lepton flavour universality would be a clear sign of physics beyond the SM, future experimental test of ratios similar to \eqref{eq:RK} are of utmost importance
\cite{Altmannshofer:2014rta,Hiller:2014ula}.

Sensitivity to interference between different short-distance contributions in \decay{\B}{\Kstar\mumu} and \decay{\Bs}{\phi\mumu} decays make angular observables powerful tests of the SM.   Combining \Bz and \Bzb decays, the angular distribution of the \decay{\Bz}{\Kstarz\mumu} decay can be described by~\cite{Altmannshofer:2008dz} 

\begin{equation}
\frac{1}{\deriv\Gamma/\deriv\qsq} \cdot \frac{\deriv^{4}\Gamma[\Bzb + \Bz]}{\deriv\cos\theta_{\ell}\,\deriv\cos\theta_{K}\,\deriv\phi\,\deriv\qsq} = \frac{9}{32\pi} \sum_{i} S_{i}(\qsq) f_{i}(\cos\theta_{\ell},\cos\theta_{K},\phi)~,
\end{equation} 

\noindent where the $S_i$ are observables that depend both on the underlying short-distance physics involved in the decay and on the $B\to K^*$ form-factors. Additional observables can be formed from the $S_i$ that exploit the cancellation of the soft form-factors, for example~\cite{Kruger:2005ep,Descotes-Genon:2013vna}

\begin{equation} 
A_{\rm T}^{2} =  P_{1} = \frac{2 S_{3}}{1 - F_{\rm L}} ~\text{and}~ P'_{4,5} = \frac{S_{4,5}}{\sqrt{F_{\rm L}( 1 - F_{\rm L})}} ~.
\end{equation}

\begin{figure}[htb]
\centering
\includegraphics[width=0.48\textwidth]{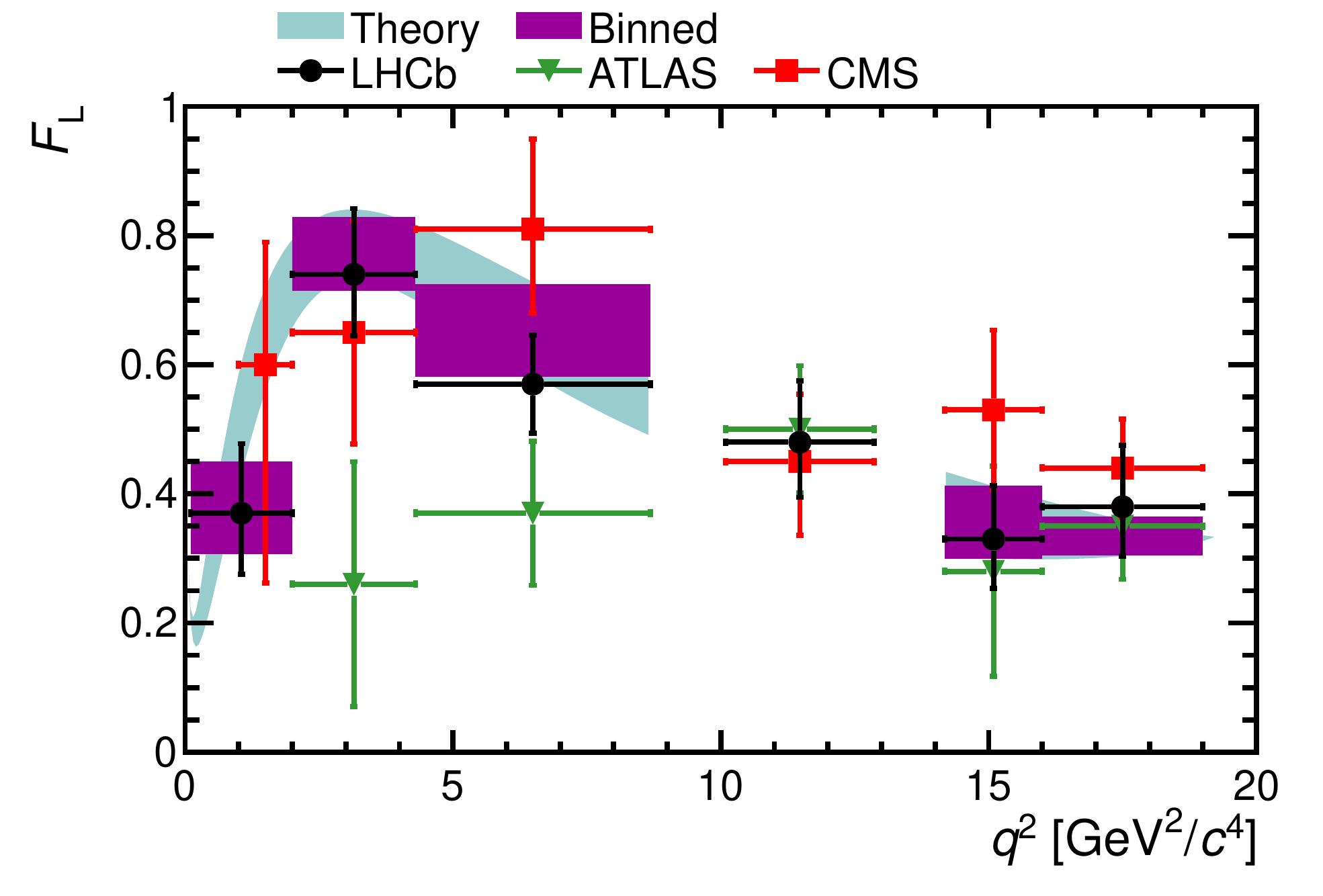} 
\includegraphics[width=0.48\textwidth]{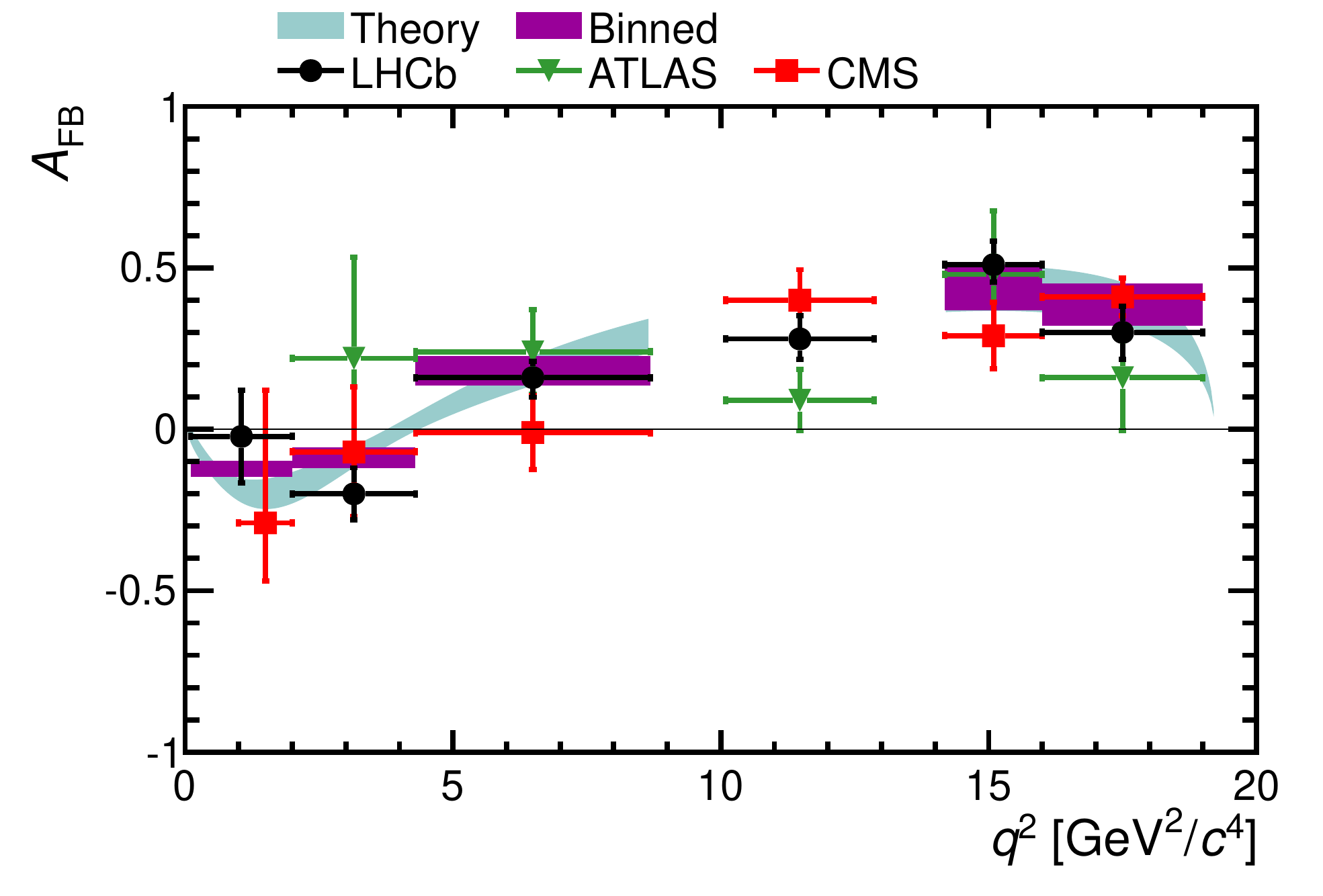}  \\ 
\caption{Longitudinal polarisation, $F_{\rm L}$, of the \Kstarz  in \decay{\Bz}{\Kstarz\mumu} decays (left) and the forward-backward asymmetry of the dimuon system, $A_{\rm FB}$ (right). Results from the ATLAS~\cite{ATLAS:KstarMuMu}, CMS~\cite{Chatrchyan:2013cda} and LHCb~\cite{LHCb-PAPER-2013-019} experiments are included. The data are overlaid with a SM prediction based on the calculation described in Ref.~\cite{Bobeth:2011gi}.} 
\label{fig:angular}
\end{figure}

\begin{figure}[htb]
\centering
\includegraphics[width=0.48\textwidth]{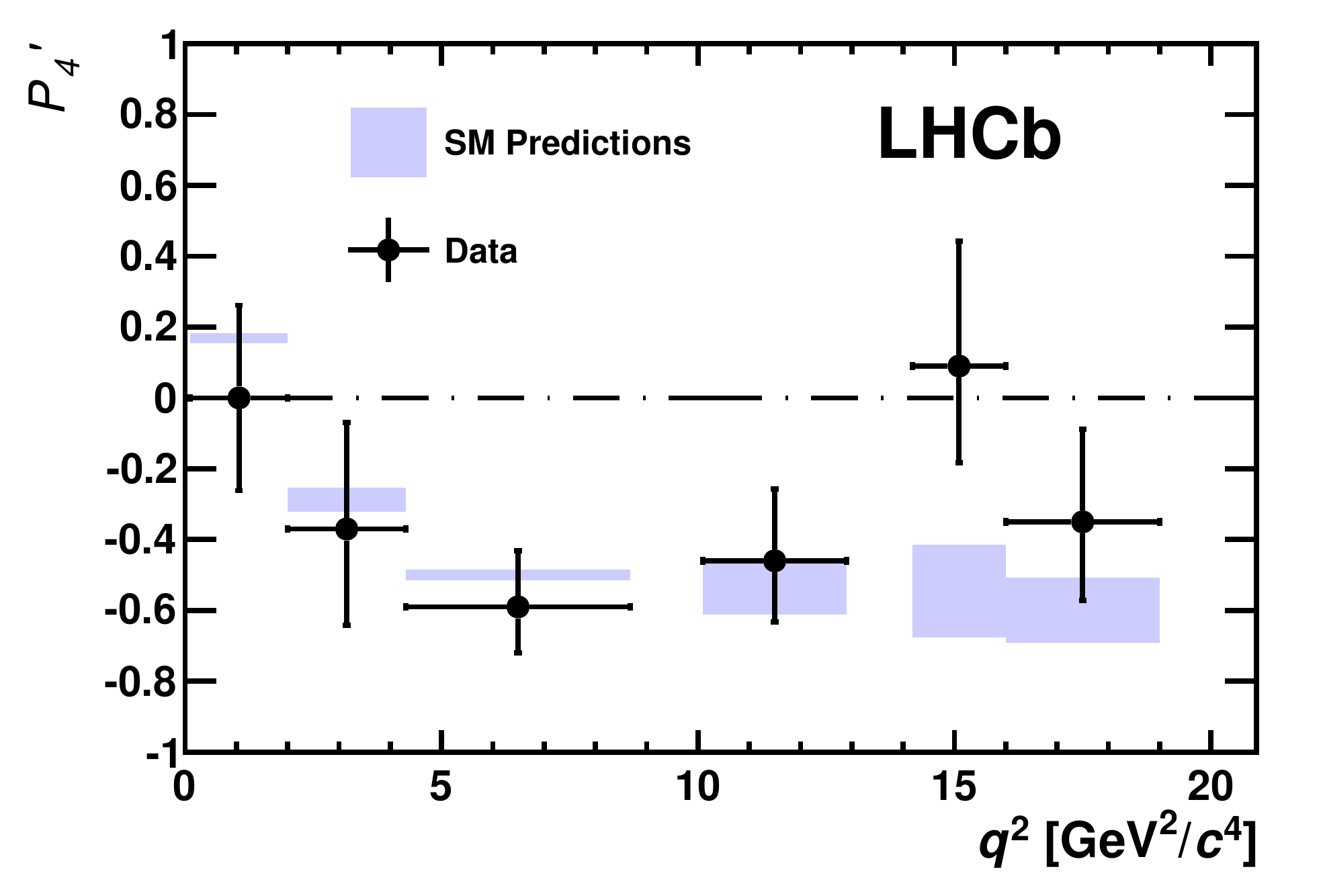} 
\includegraphics[width=0.48\textwidth]{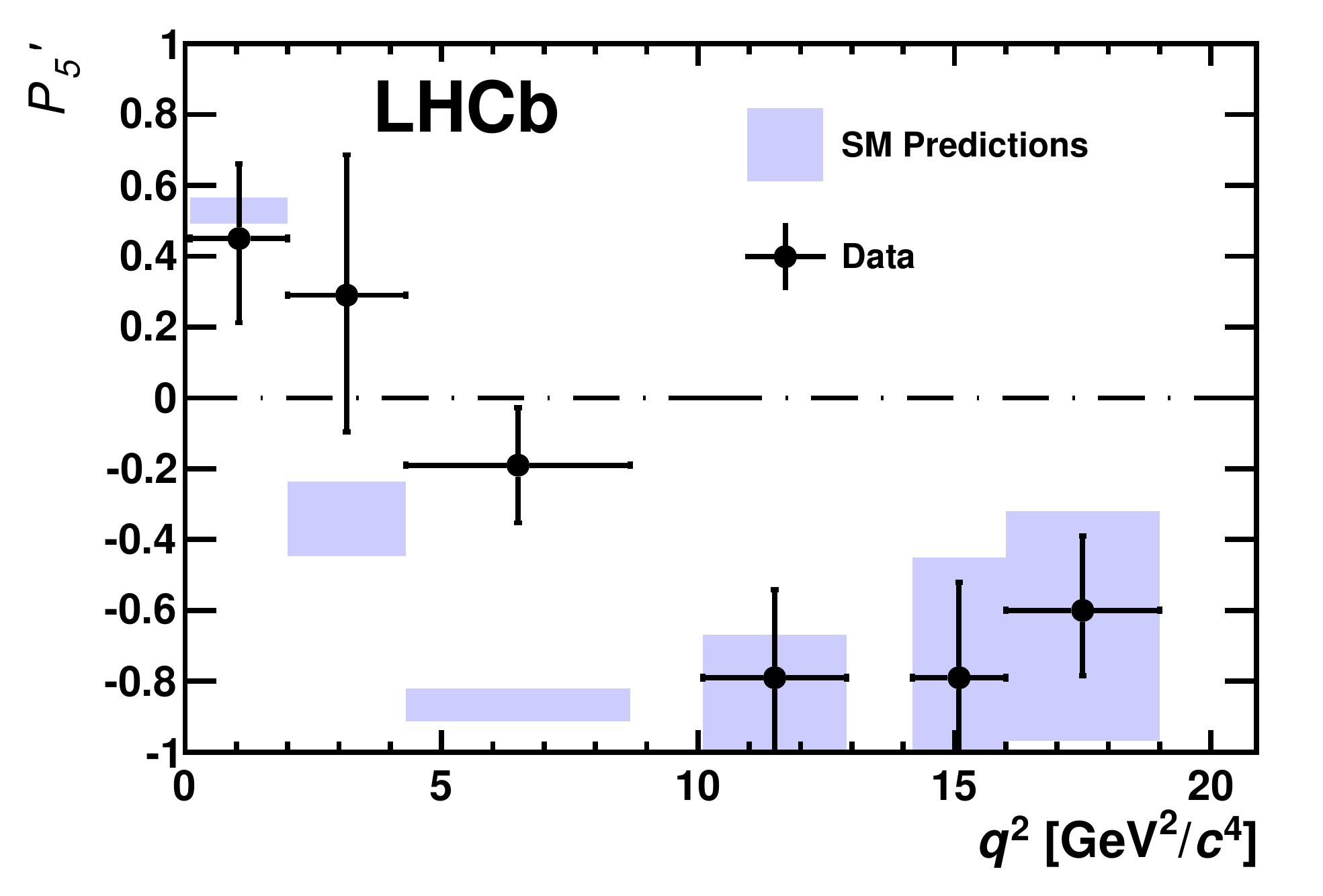}  \\ 
\caption{Observables $P'_4$ and $P'_5$, measured by the LHCb collaboration~\cite{LHCb-PAPER-2013-037} in \decay{\Bz}{\Kstarz\mumu} decays. The data are overlaid with a SM prediction described in Ref.~\cite{Descotes-Genon:2013vna}.} 
\label{fig:p5prime}
\end{figure}

The ATLAS, CMS and \lhcb collaborations have all performed measurements of the \decay{\Bz}{\Kstarz\mumu} angular distribution using the data they collected in 2011~\cite{ATLAS:KstarMuMu,Chatrchyan:2013cda,LHCb-PAPER-2013-019}. With this dataset it was not possible to perform a full angular analysis. Instead the experiments performed a partial angular analysis, using simpler projections of the angular distributions.  These projections give sensitivity to the longitudinal polarisation of the \Kstarz, $F_{\rm L}$, the forward backward asymmetry of the dimuon system (see Fig.~\ref{fig:angular}) and, in the case of LHCb, the transverse asymmetry $A_{\rm T}^{2}$.  The \lhcb experiment also performed a separate angular analysis, exploiting transformations of the angular distribution, to measure $P'_4$ and $P'_5$ (see Fig.~\ref{fig:p5prime}).  The \lhcb data for $P'_5$ shows a large local discrepancy with respect to the SM prediction at low \qsq. The other observables are consistent with the SM. 

An important open issue in exclusive \decay{b}{s\ellell} processes is the presence of broad charmonium resonances at high $q^2$. On one hand, it has been argued that above the narrow charmonimum resonances, resonance effects cancel in observables when integrating over a sufficiently large $q^2$ range~\cite{Beylich:2011aq}. On the other hand, it was shown recently that the resonance spectrum measured by LHCb in \decay{\Bp}{\Kp\mumu} decays~\cite{LHCb-PAPER-2013-039} is not well described by the pattern seen in the $R$-ratio in $e^+e^-$ collisions~\cite{Lyon:2014hpa}.

\medskip
Being free from such long distance effects, the FCNC $b \to s (d) \nu \bar{\nu}$ processes are ideal tools to search for new physics. The precision of the branching fraction predictions in the SM~\cite{Buras:2014fpa}
\begin{displaymath}
\begin{split}
\BF(B \to K^{*+} \nu \bar{\nu})_{\rm SM} &=  (9.2\pm1.0) \times 10^{-6} ~\text{and} \\
\BF(B \to K^{+} \nu \bar{\nu})_{\rm SM} &= (4.0\pm0.5) \times 10^{-6}~.\\
\end{split}
\end{displaymath}
has recently improved thanks to new results on the relevant form factors from lattice and light cone sum rules, as mentioned above. Although in certain new physics models, the precise measurements of $b\to s\ell^+\ell^-$ limit the possible deviations from the SM in $b \to s \nu \bar{\nu}$ decays, this is not true in general. For instance, new physics coupling exclusively to third generation leptons is very weakly constrained at present and could lead to a strong enhancement of these modes.

The \babar and Belle experiments have both searched for $B \to K^{(*)} \nu \bar{\nu}$ decays using their full data sets. Since there are two undetected neutrinos in the final state, the other $B$ meson from the $\Upsilon(4S)$ needs to be tagged hadronically or semi-leptonically. 
Belle reconstruct 1104 exclusive hadronic $B$ decays,
while \babar use 680 exclusive hadronic decays as tagging modes.
The results are consistent with the background only hypothesis and upper limits are set on the branching fraction of the $B \to K^{(*)} \nu \bar{\nu}$ decays~\cite{Lutz:2013ftz, Lees:2013kla}. These upper limits are still one order of magnitude larger than SM predictions, but it is expected that the Belle II experiment will be able to observe the signal at the SM branching fraction.

\subsubsection*{Inclusive semileptonic decays}

Inclusive rare semi-leptonic decays
% are important probes of new physics. They 
are complementary to the exclusive  $b \to s \ell^+\ell^-$ decays because they have a different dependence on the Wilson coefficients and rely on different theoretical methods. The latter is particularly important when disentangling new physics effects from hadronic effects in the exclusive decays.

The \babar experiment has measured the branching fraction and direct \CP asymmetry of $B \to X_s \ell^+ \ell^-$ decays using a semi-inclusive approach, where they reconstruct ten $X_s$ hadronic final states with the mass of the $X_s$ system, $M_{X_s} < 1.8\gevcc$. This covers about 70\% of the inclusive rate. The measured branching fraction is~\cite{Lees:2013nxa}
\begin{displaymath}
\BF(B \to X_s \ell^+ \ell^-) = (6.73^{\,+0.70}_{\,-0.64}{}^{\,+0.34}_{\,-0.25} \pm 0.50) \times 10^{-6} ~,
\end{displaymath}
which is consistent with the corresponding SM prediction~\cite{Huber:2007vv}. They also extract the direct \CP asymmetry in $B \to X_s \ell^+ \ell^-$ decays using same data set with seven $X_s$ self-flavour-tagged hadronic final states, yielding
\begin{displaymath}
\mathcal{A}_{\CP}(B \to X_s \ell^+ \ell^-) = 0.04 \pm 0.11 \pm 0.01~,
\end{displaymath}
which is consistent with a null asymmetry, expected in the SM.

\begin{figure}[!htb]
\centering
\includegraphics[width=0.8\linewidth]{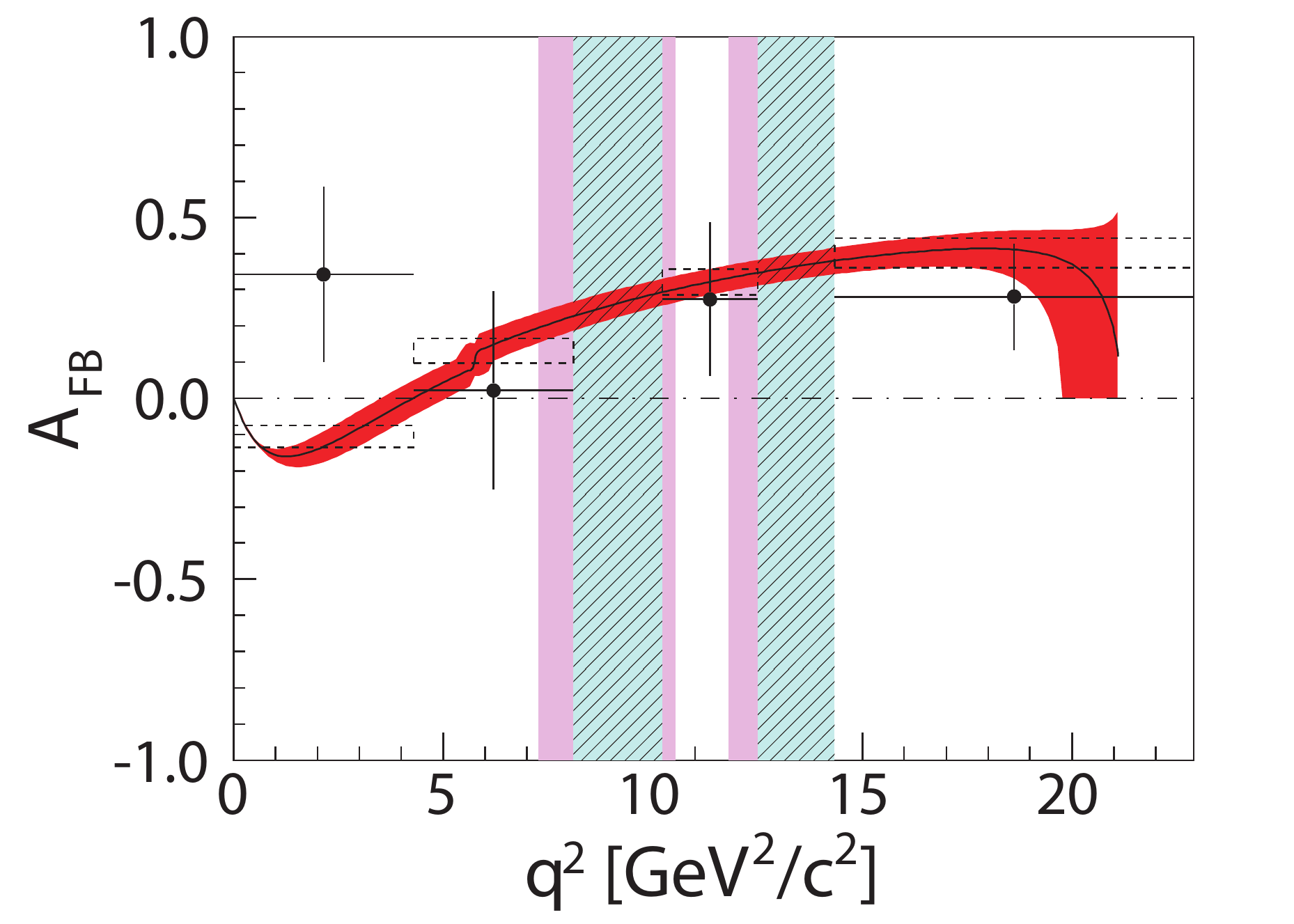}
\caption{
Forward-backward asymmetry, $A_{\rm FB}$, of inclusive $B \to X_s \ell^+ \ell^-$ decays measured by the Belle experiment in bins of dimuon invariant mass squared, \qsq. The continuous band (red) indicates the SM prediction. The dashed open boxes represent the SM prediction, averaged over the same \qsq interval as the data points. Reproduced from Ref.~\cite{Sato:2014pjr}.
\label{fig:belle14:afbbsll}
}
\end{figure}

The Belle experiment has recently measured the forward-backward asymmetry of $B \to X_s \ell^+ \ell^-$ decays using a similar semi-inclusive technique~\cite{Sato:2014pjr}. They reconstruct 18 $X_s$ hadronic final states with a $M_{X_s} < 2.0~\gevcc$ selection. Belle use ten flavour specific modes, which corresponds to 50\% of the total inclusive rate, to measure the forward-backward asymmetry of the dilepton system. The measured asymmetry is summarised in Fig.~\ref{fig:belle14:afbbsll} and is consistent with the SM~\cite{Ali:2002jg}. This is the first measurement of the forward-backward asymmetry in inclusive $B \to X_s \ell^+ \ell^-$ decays.

\subsection{Radiative decays} 

The exclusive and inclusive rare decays based on the $b\to s\gamma$ transition are important processes to probe new physics in the electromagnetic dipole operators, $Q_{7}$ and $Q_7^{\prime}$ (as well as as the chromomagnetic dipole operators entering through renormalisation group mixing and through sub-leading hadronic contributions in the exclusive decays). While the sum of squared absolute values of the Wilson coefficients $C_7$ and $C_7'$ is strongly constrained to be close to its SM value by the measurement of the inclusive $B\to X_s\gamma$ branching ratio, two important open questions are a possible non-standard \CP-violating phase as well as the photon polarisation, i.e.\ the relative strength of $C_7$ and $C_7'$. Photons emitted in $b \to s \gamma$ process are predominantly left handed
in the SM while new physics with right handed current allows large right handed polarisation.
In addition to $B\to K^{*} \ell^+\ell^-$ angular observables discussed in section~\ref{sec:sl}, two important observables to access the photon polarisation are the mixing-induced \CP asymmetry, $S_{K^*\gamma}$, in $B^0\to K^{*0}\gamma$ decays and the up-down asymmetry $\lambda_\gamma$ in $B\to K_1(\to \Kp\pim\pip)\gamma$ decays (see \cite{Becirevic:2012dx} and references therein). 

%The \babar and Belle experiments have measured~\cite{Aubert:2008gy,Asner:2010qj,Ushiroda:2006fi} to be $S_{\Kstar\gamma} = -0.16\pm 0.22$.  This measurement is consistent with the SM expectation of -2.6\%~\cite{Ball:2006eu}. 
\subsubsection*{Exclusive radiative decays}

%Photons emitted in $b \to s \gamma$ process are predominantly left handed since the weak interaction has ${\rm SU}(2)_{\rm L}$ gauge symmetry in the SM while new physics with right handed current allows large right handed polarisation. There are five methods proposed to measure the photon polarisation in b-hadron decays: time dependent $CP$ asymmetry in $B^0 \to f_{CP}^0 \gamma$ decays; up-down asymmetry of photons in $B \to K_1 \gamma \to K \pi \pi \gamma$ decays; angular analysis of converted photons in $b \to s \gamma$ decays; analysis of the photon direction in $b-$baryon decays; and an angular analysis of low $q^2$ region in $B \to K* \ell^+ \ell^-$ (where the photon is virtual).

The \babar experiment has measured time dependent \CP asymmetry in $B^0 \to \KS \rho^0 \gamma$. To estimate the dilution factor $D_{K_{S}\rho}$ from $B^0 \to K^{*+} \pi^- \gamma$ flavour eigenstate process with the same final state, a Dalitz plot analysis for $B^+ \to K^+ \pi^- \pi^+$ was also performed. The $CP$ violation parameter $S_{K_{S}\pi \pi}$ and dilution factor were measured as 
\begin{displaymath}
\begin{split}
S_{K_{S}\pi \pi} & = 0.14 \pm 0.25 {}^{+0.04}_{-0.03} ~~\text{and}\\ 
D_{K_{S}\rho} & = 0.549 {}^{+0.096}_{-0.004} \\ 
\end{split}
\end{displaymath}
%$S_{K_{S}\pi \pi} = 0.14 \pm 0.25 {}^{+0.04}_{0.03}$ and $D_{K_{S}\rho} = 0.549 {}^{+0.096}_{-0.004}$. 
From these two values, the \CP violation parameter is obtained as
\begin{displaymath}
S_{K_{S}\rho} = 0.25 \pm 0.46 {}^{+0.08}_{-0.06} ~,
\end{displaymath}
which is consistent with null asymmetry. 

\begin{figure}[!htb]
\centering
\includegraphics[width=0.8\linewidth]{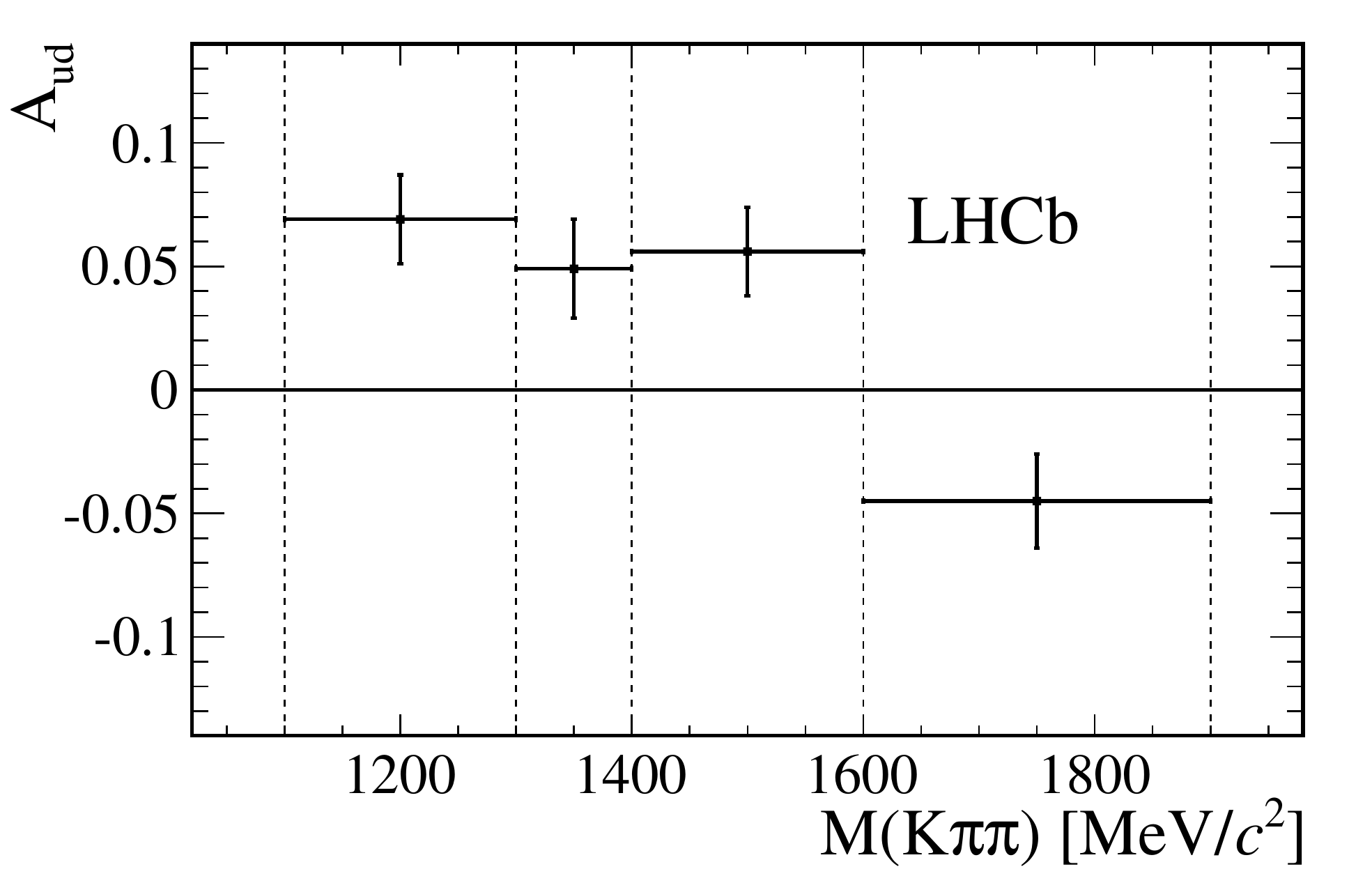} 
\caption{
Up-down asymmetry, $A_{\rm ud}$, of the photon direction with respect to the $\Kp\pim\pip$ system in $B^+ \to K^+ \pi^- \pi^+ \gamma$ decays, measured in four bins of $\Kp\pim\pip$ invariant mass. Reproduced from Ref.~\cite{LHCb-PAPER-2014-001}.
\label{fig:lhcb:aud}
}
\end{figure}

The LHCb experiment searched for evidence of photon polarisation using the up-down asymmetry of photons in $B^+ \to K^+ \pi^- \pi^+ \gamma$ processes. From the reconstructed $K^+ \pi^- \pi^+$ mass of the candidates it is evident that not only $1^+$ resonances but also  $1^-$, $2^+$ and $2^-$ resonances might be present in the $\Kp\pim\pip$ system. The up-down asymmetry is studied inclusively in four $M_{K\pi\pi}$ mass bins. The measured up-down asymmetry is summarised in Fig.~\ref{fig:lhcb:aud} and the total significance is 5.2$\sigma$ for non-zero asymmetry~\cite{LHCb-PAPER-2014-001}. This is the first observation of photon polarisation in radiative $B$ decays. However, a quantitative interpretation of the up-down asymmetry in terms of $C_7$ and $C'_7$ needs a deeper understanding of the admixture of resonances that contribute to the $K^+ \pi^- \pi^+ \gamma$ final state. 
%to photon polarisation from $A_{\rm UD}$ is on-going.

\subsubsection*{Inclusive radiative decays}

The Belle experiment presently provides the best measurement of the branching fraction of $B \to X_s \gamma$, which is sensitive to the absolute value of Wilson coefficients $|C_7|^2 + |C_7'|^2$. The hadronic system $X_s$ is reconstructed with sum-of-exclusive technique using 38 final states with a hadronic mass less than 2.8~GeV/$c^2$ which corresponds to photon energy of 1.9\gev. The measured branching fraction is 
\begin{displaymath}
\BF(B \to X_s \gamma; M_{X_s} < 2.8\gev) =(3.51 \pm 0.17 \pm 0.33 ) \times 10^{-4}
\end{displaymath}
which is extrapolated to photon energy of 1.6~GeV to be~\cite{Saito:2014das}.
\begin{displaymath}
\BF(B \to X_s \gamma; E_{\gamma} > 1.6~{\rm GeV}) =(3.74 \pm 0.18 \pm 0.35 ) \times 10^{-4}~.
\end{displaymath}
This is the world most precise measurement with a sum-of-exclusive technique and is consistent with theoretical predictions~\cite{Misiak:2006zs, Becher:2006pu},
%\begin{displaymath}
%\BF(B \to X_s \gamma)_{\rm SM} =(3.15 \pm 0.23 ) \times 10^{-4} 
%\end{displaymath} 
%\BF(B \to X_s \gamma) =(2.98 \pm 0.26 ) \times 10^{-4}$~\cite{Becher:2006pu}.

Direct \CP asymmetries in $b \to s (d) \gamma$ can be used to probe imaginary parts of $C_7$ and $C_7'$. Theoretical prediction of the \CP asymmetries have rather large uncertainties that cancel when summing $b \to s \gamma$ and $b \to d \gamma$. The \CP asymmetry of he sum is zero with  negligibly  uncertainty due to unitarity of the CKM matrix. A difference of \CP asymmetries, $\Delta \mathcal{A}_{\CP}$, between $B^0$ and $B^+$ is sensitive to an imaginary part of $C_8/C_7$ where $C_8$ is a chromo-magnetic coefficient \cite{Benzke:2010tq}. The \babar experiment reconstructs $B \to X_s \gamma$ with a sum-of-exclusive technique using 38 $X_s$ final states in which 16 self-flavour-tagged final states were used for the measurements of \CP asymmetry and difference of \CP asymmetries. The hadronic mass was required less than 2.0~GeV$/c^2$ which corresponds to photon energy greater than 2.3~GeV. The results~\cite{Lees:2014uoa}
\begin{displaymath}
\begin{split}
\mathcal{A}_{\CP} (B \to X_s \gamma) &= (1.7 \pm 1.9 \pm 1.0) \times 10^{-2}~~\text{and}  \\
\Delta \mathcal{A}_{\CP} (B \to X_s \gamma) &= (5.0 \pm 3.9 \pm 1.5) \times 10^{-2} \\
\end{split}
\end{displaymath}
are consistent with zero and theoretical predictions in the SM. From the $\Delta \mathcal{A}_{\CP}$, a limit on $\Imag(C_8/C_7)$ was also set for the first time. 
The Belle experiment searched for the \CP asymmetry in a sum of $b \to s \gamma$ and $b \to d \gamma$ with a fully inclusive reconstruction technique which only measure a prompt photon with energy greater than 2.1\gev for signal $B$ meson, and a lepton with momentum between 1.1\gev and 2.25\gev from the other $B$ meson to tag the flavour of the signal $B$ meson. The measured raw asymmetry is corrected for a dilution arising from \Bz-\Bzb mixing and wrong tagging due to secondary leptons and misidentifications of hadrons as leptons. The Belle experiment measures 
\begin{displaymath}
\mathcal{A}_{\CP} ( B \to X_{s+d} \gamma ) = (2.2 \pm 4.0 \pm 0.8) \times 10^{-2}~,
\end{displaymath}
which again is consistent with null asymmetry.

\subsection{Hadronic decays}

Rare charmless hadronic $B$ decays are useful both to probe CKM elements and to probe physics beyond the Standard Model. Theoretically, they are more challenging than semi-leptonic decays. Getting a handle on the strong amplitudes can be either achieved in a data-driven way by making use of approximate flavour symmetries, or by computing the amplitudes within a framework such as QCD factorisation (QCDF), soft-collinear effective theory (SCET), or perturbative QCD (pQCD).
Large numbers of experimental measurements can also be leveraged to constrain new physics through global analyses 
(see e.g.~\cite{Bobeth:2014rra} for a recent example).

Baryonic $B$-meson decays are one of the least known decay processes; since there are many contributing diagrams and final state quarks, which make theoretical calculation difficult. Naive predictions suggest that two-body baryonic decays have smaller branching fractions than those of three-body decays. LHCb observed a two-body baryonic $B$ decay $B^+ \to \overline{\Lambda}(1520) p$ and found an evidence of $B^0 \to p \overline{p}$ with branching fractions of~\cite{LHCb-PAPER-2013-031}
\begin{displaymath}
\begin{split}
\BF(\decay{\Bp}{\overline{\Lambda}(1520)p}) &= (3.9^{+1.0}_{-0.9}\pm{0.1}\pm{0.3}) \times 10^{-7} ~\text{and} \\ 
\BF(\decay{\Bp}{p\overline{p}}) &= (1.5^{+0.6}_{-0.5}{}^{+0.4}_{-0.1}) \times 10^{-8} 
\end{split}
\end{displaymath} 
which appears to confirm the naive prediction. Three-body baryonic $B^+ \to p \overline{p} h^+$ decays were also studied by LHCb. The branching fractions and threshold enhancements observed by Belle~\cite{Chang:2008yw,*Wei:2007fg,*Chen:2008jy,*Wang:2007as,*Wang:2005fc,*Lee:2004mg} were confirmed and \CP violation in $B \to p \bar{p} K^+$ for proton-kaon system  mass squared larger than $10\gev^{2}$ was found with $4\sigma$ significance~\cite{LHCb-PAPER-2014-034}.

Three-body charmless $B^+ \to h^+h^-h^+$ decays proceed via $b \to s(d)$ penguin and/or $b \to u$ tree diagrams.  Interferences between the different diagrams diagrams and between different intermediate resonances can lead to large \CP violation in the decay. LHCb searched for, and observed, large \CP violation in $B^+ \to K^+ K^- K^+$, $B^+ \to K^+ \pi^- \pi^+$, $B^+ \to \pi^+ \pi^- \pi^+$ and $B^+ \to \pi^+ K^- K^+$ decays. For example,
\begin{displaymath} 
\mathcal{A}_{\CP}(B^+ \to \pi^+ K^- K^+) = -0.123 \pm 0.017 \pm 0.012 \pm 0.007
\end{displaymath}
in $B^+ \to \pi^+ K^- K^+$ decays~\cite{LHCb-PAPER-2014-044}.

\subsection{Global analyses of $b \to s$ data}

\begin{figure}[tb]
\includegraphics[width=0.45\textwidth]{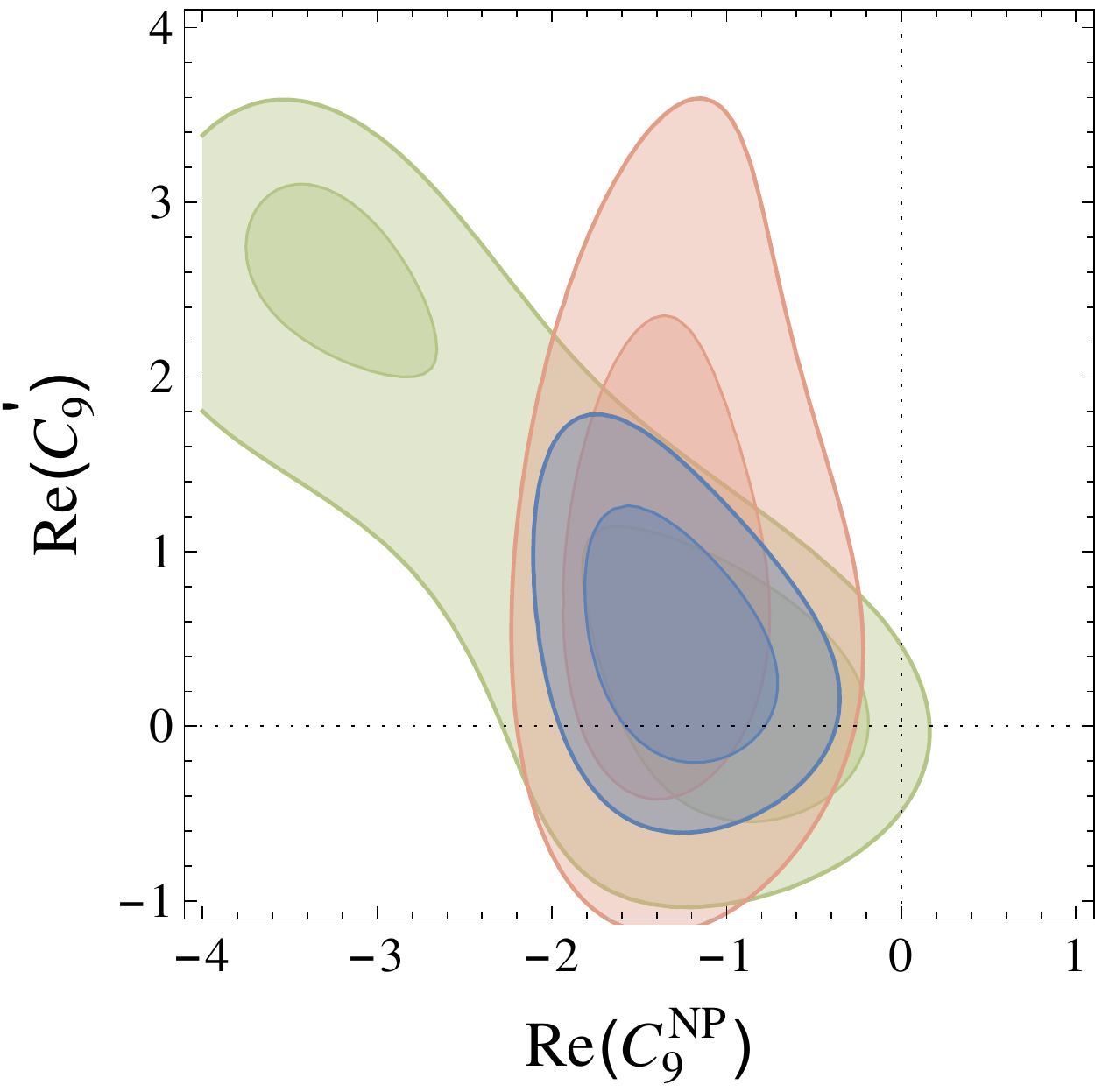}%
\hspace{0.05\textwidth}%
\includegraphics[width=0.45\textwidth]{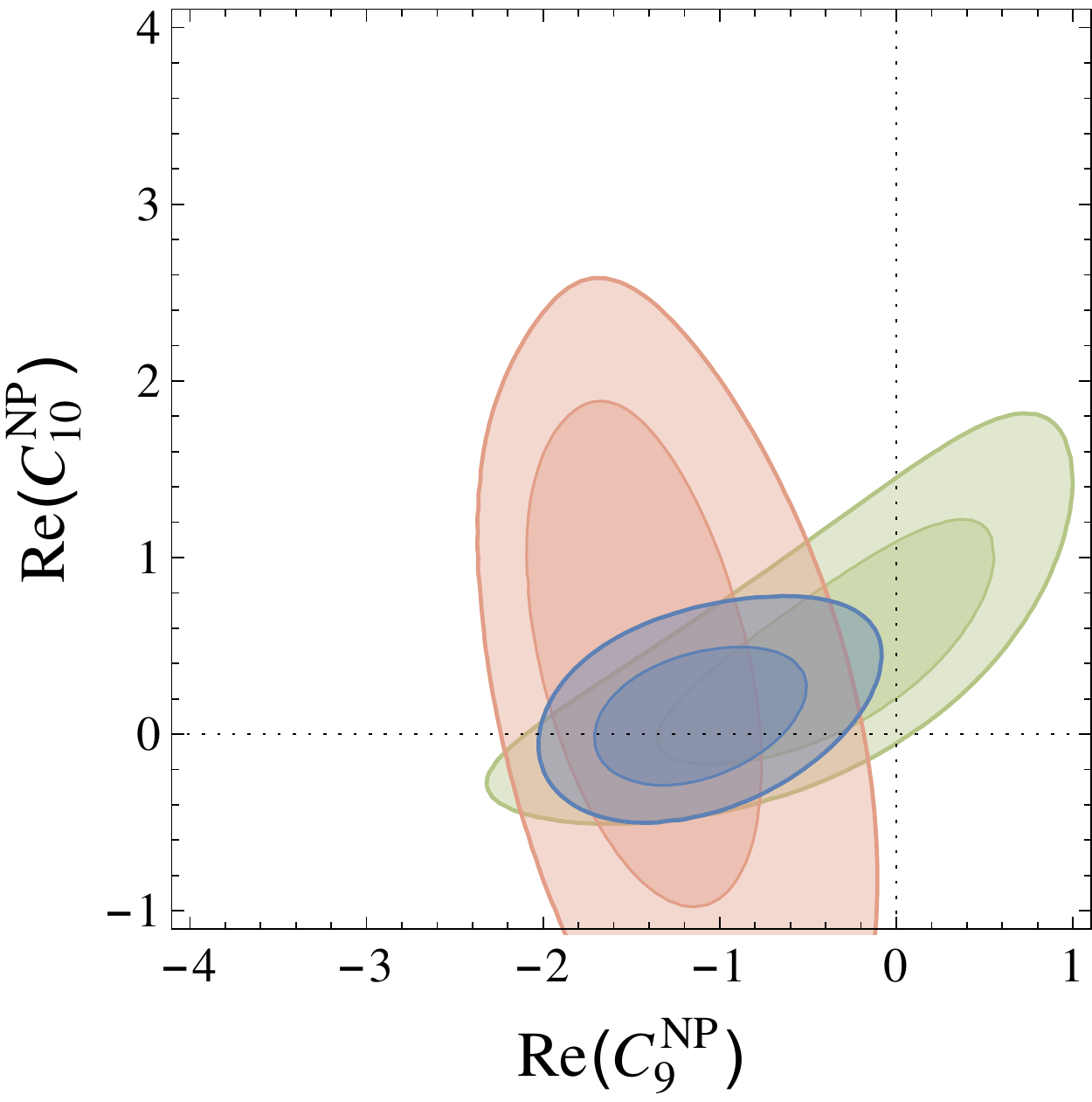}%
\caption{Allowed regions in the the plane of new physics contributions to the Wilson coefficients $C_9$, $C_9'$ and $C_{10}$ from a global fit (blue), using only branching ratio data (red), or only data on $B \to K^* \mu^+\mu^-$ angular observables (green). Taken from \cite{Altmannshofer:2014rta}.}
\label{fig:obs}
\end{figure}

The measurements of leptonic, semi-leptonic, and radiative $B$ meson decays can be used for a model-independent analysis of new physics effects in $b\to s$ transitions. A global fit of new physics contributions to the Wilson coefficients in the effective Hamiltonian
(\ref{eq:Heff})
finds a preference for a negative new physics contribution to the Wilson coefficient $C_9$ \cite{Altmannshofer:2014rta} (cf.~also the earlier work \cite{Descotes-Genon:2013wba,Altmannshofer:2013foa,Beaujean:2013soa,Hurth:2013ssa}), possibly with a positive contribution to the Wilson coefficients $C_9'$ or $C_{10}$. An import role is played both by $B\to K^*\mu^+\mu^-$ angular observables and by branching ratio measurements of $B\to K^*\mu^+\mu^-$, $B\to K\mu^+\mu^-$, and $B_s\to \phi\mu^+\mu^-$ (see fig.~\ref{fig:obs}). The branching fraction of these exclusive decay measured by the LHCb experiment is systematically below that predicted in the SM~\cite{LHCB-PAPER-2014-006}.
An interesting model that can explain such an effect is a $Z'$ model with gauged muon minus tau number \cite{Altmannshofer:2014cfa}. This model would also predict $R_{\rm K} \approx 0.75$, close to what is measured in data. Other $Z'$ models have been suggested in the literature as well \cite{Gauld:2013qba,Buras:2013qja,Gauld:2013qja,Buras:2013dea}. The Wilson coefficients $C_7$ and $C'_7$, responsible for radiative decays, are found to be consistent with the SM expectation.

Future improved measurements of $B\to K^*\mu^+\mu^-$ angular observables will allow for a significantly more precise determination of the Wilson coefficients. In this respect, an important role is played by angular CP asymmetries. These will allow to determine the imaginary parts of the Wilson coefficients, which are still constrained very weakly.

\section{Rare charm decays}

Rare decays of  $D$-meson allow to probe flavour-changing neutral currents involving up-type quarks and are therefore of high theoretical interest. In the SM, the short-distance contribution is strongly GIM suppressed and long-distance effects are expected to dominate. This makes disentangling a potential new physics contribution from the Standard Model background very challenging.

Of particular interest are contributions from the magnetic and chromomagnetic dipole operators, which are generated in many well-motivated new physics models, e.g.\ the MSSM \cite{Prelovsek:2000xy}, Randall-Sundrum models \cite{Delaunay:2012cz}, or composite Higgs models \cite{Konig:2014iqa}. Such effects could also explain a possible deviation from the SM in the direct \CP asymmetry difference $\Delta \mathcal{A}_{\CP}$ between $\Dz\to \pip\pim$ and $\Dz\to \Kp\Km$ decays. An important cross-check of new physics in charmed dipole operators is given by \CP asymmetries in $D\to V\gamma$ decays \cite{Isidori:2012yx,Lyon:2012fk}.

In the case of semileptonic decays, $D\to (X_u,P,V)\to \ell^+\ell^-$, considering \CP or angular asymmetries can also be helpful in disentangling short-distance from long-distance contributions. The purely leptonic decay $\Dz\to\mu^+\mu^-$ is a special case since the dominant long-distance contribution could be determined from data by measuring the $\Dz\to \gamma\gamma$ decay \cite{Burdman:2001tf}, allowing to access a possible new physics contribution.

The \babar experiment has measured branching fractions of radiative $D^0 \to \phi \gamma$ and $D^0 \to \bar{K}^{*0} \gamma$ decays. Signals are extracted by two-dimensional fit to $D^0$ mass and helicity angle distribution of vector mesons. The branching fractions are measured to be~\cite{Aubert:2008ai}
\begin{displaymath}
\begin{split}
\BF(D^0 \to \phi \gamma) & = (2.73 \pm 0.30 \pm 0.26) \times 10^{-5} ~\text{and} \\
\BF(D^0 \to \bar{K}^{*0} \gamma) & = (3.22 \pm 0.20 \pm 0.27) \times 10^{-4}
\end{split}
\end{displaymath}
\babar also searched for a photonic decay, $D^0 \to \gamma \gamma$ that branching fraction is predicted to be ${\cal{O}}(10^{-8})$ in the SM while up to 200 times enhancement is possible in the MSSM. \babar set the best limit 
\begin{displaymath}
\BF(D^0 \to \gamma \gamma) < 2.2 \times 10^{-6}~\text{at~90\%~C.L.}
\end{displaymath}
which already probes the new physics parameter space~\cite{Lees:2011qz}.

Leptonic $D$ decays are helicity and loop suppressed, thus important tool to search for scalar particle contributions. LHCb searched for $D^0 \to \mu^+ \mu^-$ channel using $D^{0*}$ tagging. An important background is $D^0 \to \pi^+ \pi^-$ decays where both pions are misidentified as muons. The probability of a pion misidentified as a muon is estimated using $D^0 \to K^+ \pi^-$ samples. The upper limit obtained is~\cite{LHCb-PAPER-2013-013} 
\begin{displaymath}
\BF(D^0 \to \mu^+ \mu^-) < 6.2 \times 10^{-9}~,
\end{displaymath}
which is a still at least two orders of magnitude from the SM prediction (which is dominated by long-distance effects).% Belle studied modes involving electrons and gave the best upper limits, $\BF (D^0 \to e^+ e^-) < 7.9 \times 10^{-8}$ and $\BF(D^0 \to e^{\pm} \mu^{\mp}) < 2.6 \times 10^{-7}$. 

$D$ decays to dimuon and pion(s) proceeds through different diagrams, $D^+ \to \pi^+ \mu^+ \mu^-$ and $D^0 \to \pi^+ \pi^- \mu^+ \mu^-$ via penguin or box diagrams, $D^+_{s} \to \pi^+ \mu^+ \mu^-$ via a weak annihilation diagram. The LHCb experiment searched for these decays and has improved the upper limits on these decays by more than one order of magnitude~\cite{LHCb-PAPER-2013-050}.

\section{Rare strange decays} 

The theoretically cleanest kaon decays sensitive to new physics are the $K^+\to\pi^+\nu\bar\nu$ and $\KL \to\pi^0\nu\bar{\nu}$ decays. Their branching ratios are predicted in the Standard Model to be \cite{Brod:2010hi}
\begin{displaymath}
\begin{split}
\BF(K^+\to\pi^+\nu\bar\nu)_{\rm SM} &= (7.81\pm0.80)\times 10^{-11}\,,
 \\
\BF(\KL\to\pi^0\nu\bar\nu)_{\rm SM} &= (2.43\pm0.39)\times 10^{-11}\,.
\end{split}
\end{displaymath}
In both cases, the dominant uncertainties are due to CKM elements.

While the semi-leptonic decays $\KL \to \pi^0\ell^+\ell^-$, with $\ell=e$ or $\mu$, are less clean theoretically, their measurement would also be an important test of of new physics since they probe short-distance physics complementary to the one probed by $K\to\pi\nu\bar\nu$ (see e.g.~\cite{Buchalla:2003sj,Mescia:2006jd,Mertens:2011ts}).

The KOTO experiment at J-PARC is a successor to the E391a experiment and is designed to observe $\KL \to \pi^0 \nu^0 \bar{\nu}^0$ decays using in-flight decays of the \KL meson. The detector was installed in 2013 and first physics run were taken during May that year. Results based on 100 hours of data taking at 10\% of design intensity were presented for the first time at CKM 2014. The performance of CsI calorimeter was checked by six photons from $\KL \to 3 \pi^0$ decays and found that the invariant mass, z vertex position distributions and photon veto performance are consistent with MC expectations. The signal events are identified with $p_T$ and $z$ vertex of the $\pi^0$. After applying loose selections, numbers of events in outside the signal box are consistent with backgrounds estimated with MC and special hadron interaction runs. The single event sensitivity of the first physics run is $1.29 \times 10^{-8}$, which is almost same as that at E391a, $1.11 \times 10^{-8}$. After opening the signal box, one 
event was observed which is consistent with backgrounds, $0.36 \pm 0.16$. The next run is be planned for early 2015. Control over background will be important if the KOTO experiment is to reach a sensitivity at the level of the SM branching fraction.

The NA62 experiments at the CERN SPS is designed to collect ${\cal{O}}(100)$ $K^+ \to \pi^+ \nu \bar{\nu}$ events in a three year run. Former $K^+ \to \pi^+ \nu \bar{\nu}$ experiments, E787/E949 at Brookhaven, measured a branching fraction
\begin{displaymath}
\BF(K^+ \to \pi^+ \nu \bar{\nu}) =1.73^{+1.15}_{-1.05} \times 10^{-10}~,
\end{displaymath}
using stopped $K^+$. The NA62 uses in-flight $K^+$. From the precisely measured kaon and pion momenta, missing mass squared is calculated and is used to discriminate signal from backgrounds. Kaon position, momentum and timing are measured by a hybrid silicon pixel detector and a differential Cherenkov counter filled with ${\rm H}_2$ or ${\rm N}_2$, which gives 95\% kaon efficiency and sub-percent pion misidentification. Pions from kaon decays are reconstructed with ultra-light straw tubes and RICH counter filled with Neon. A large background from $K^+ \to \mu^+ \nu$ whose branching fraction is 63\%, is suppressed by liquid krypton (LKr) and iron-scintillator calorimeters. Photons from $K^+ \to \pi^+ \pi^0$, whose branching fraction is 21\%, is suppressed by three calorimeters, which cover different polar angle regions. The expected performance of the NA62 detector is 45 signal events/year with less than 10 background events expected. A pilot physics run with 60 days of beam 
started in October 2014 for commissioning of hardware with a low intensity beam. A full physics run with 100 days beam time is scheduled in 2015-2017.

\section{Null tests of the SM} 

In the SM, global lepton number is conserved to an excellent precision and processes violating individual (charged) lepton family numbers are strongly suppressed by the smallness of the neutrino masses. Consequently, processes violating these symmetries are excellent null tests of the SM.
Charged lepton flavour violation can be probed in $\tau$ and $\mu$ decays, but also in rare leptonic or semi-leptonic $B$ decays. The discovery of the Higgs boson has also opened the possibility to look for lepton flavour violating Higgs decays \cite{Blankenburg:2012ex,Harnik:2012pb}. Although these decays are strongly constrained by the non-observation of radiative lepton decays in many models (see e.g.~\cite{Falkowski:2013jya}), visible effects are still possible in particular models  \cite{Kopp:2014rva}.

The MEG experiment  at PSI, which provides world's most powerful proton beam of 1.4~MW, searched for $\mu^+ \to e^+ \gamma$ decays. MEG sets an upper limit on the branching fraction of
\begin{displaymath}
\BF(\mu^+ \to e^+ \gamma) < 5.7 \times 10^{-13}
\end{displaymath}
 at 90 \% C.L.~\cite{Adam:2013mnn}. The MEG has been upgraded aiming at an eventual sensitivity of $10^{-14}$.

The HiMB experiment is a new project to search for $\mu^+ \to e^+ e^- e^+$ at PSI. The experiment aims to achieve a $10^{-15}$ upper limit in phase 1 and targets $10^{-16}$ in a phase 2 upgrade. Ultra thin trackers are required to minimise electron multiple scattering. HV-MAPS with flex print on kapton frame is a candidate tracker whose radiation length is only 0.1\% per layer.

Two $\mu-e$ conversion experiments are also planned. One is COMET at J-PARC and the other is Mu2e at Fermilab. Both experiments plan to use straw tube and calorimeter systems for electron tracking and identification.  The COMET experiment aims at the upper limit for $\mu^- + \text{Al}
 \to e^- + \text{Al}$ conversion rate of $10^{-14}$ at the first stage, and then explore down to $10^{-16}$ at phase 2 with full detector. 

LFV processes are also searched with the heaviest lepton $\tau$. At $e^+e^-$ colliders, various LFV $\tau$ decay can be searched and can be tested correlations among modes to identify new physics models. While at hadron colliders, $D_s^+ \to \tau^+ \nu_{\tau}$ is the dominant production and final states involving muons can be searched.

\babar and Belle use the almost same technique as $B$ meson reconstruction to search for $\tau^{+} \to \mu^{+} \gamma$ and $\tau^{+} \to e^{+} \gamma$ with integrated luminosities of 516\invfb and 535\invfb~\cite{Hayasaka:2007vc}, respectively. \babar gave the world's best upper limits of~\cite{Aubert:2009ag}
\begin{displaymath}
\begin{split}
\BF(\tau^+ \to \mu^+ \gamma) < 4.4 \times 10^{-8} \mbox{ and }
\BF(\tau^+ \to e^+ \gamma) < 3.3 \times 10^{-8}.
\end{split}
\end{displaymath}
An update from Belle is anticipated in the near future.

The $\tau^{+} \to \mu^{+}\mu^{-}\mu^{+}$ process is searched \babar, Belle and LHCb. At the LHCb, a displaced decay vertex is the powerful tool also for short lived $\tau$ lepton. A limit of 
\begin{displaymath}
\BF(\tau^{+} \to \mu^{+}\mu^{-}\mu^{+}) < 4.6 \times 10^{-8}
\end{displaymath}
is obtained by LHCb~\cite{LHCb-PAPER-2013-014,*LHCb-PAPER-2014-052}, which is comparable to the branching fractions limits of $3.3 \times 10^{-8}$ and $2.1 \times 10^{-8}$  set by Babar~\cite{Lees:2010ez} and Belle~\cite{Hayasaka:2010np}, respectively.

The neutrino oscillation observed by Super-Kamiokande~\cite{Fukuda:1998mi} had proved that neutrinos are massive. Next big issue in neutrino physics is determination of neutrino type, Dirac or Majorana, since Majorana neutrino with \CP violation could lead leptogenesis~\cite{Fukugita:1986hr}. If Majorana neutrinos are light, below $M_B$, lepton number violating $B^+ \to X^- \ell^+ \ell^+$ decays can be occurred by exchanging the Majorana neutrinos~\cite{Atre:2009rg}. \babar and LHCb gave new results on search for $B^+ \to (\pi^-, K^-, \rho^-, K^{*-}, D^-) \ell^+ \ell^+$ and set about one order of magnitude more stringent upper limits on the branching fractions from CKM 2012~\cite{LHCb-PAPER-2013-064,*LHCb-PAPER-2011-038,*Lees:2013gdj}.

\clearpage

\section{Summary}

There has been huge progress since the last CKM conference in 2012. The last two years have seen important results from the legacy datasets from the $B$-factories and many new results from the ATLAS, CMS and LHCb experiments from the first run of data taking at the LHC. There have also been new results from MEG experiment at PSI, new results on rare charm decays from BES III and commissioning runs for KOTO and NA62.

The next couple of years promise to be equally productive, with many new results anticipated from the LHC experiments, the final results from MEG and the first full data taking runs from KOTO and NA62. These will also be important years for the Belle II experiment as it prepares for data taking in 2017. 

The wealth of experimental data collected on rare decays are more or less consistent with SM. However, some interesting tensions have started to emerge in exclusive semileptonic $b \to s \ell^+ \ell^-$ decays.

%%%%%%%%%%%%%%%%%%%%%%%%%%%%%%%%%%%%%%%%%%%%%%%%%%%%%%%%%%%%%%%%%%%%%%%%%
%%
%%   use this format to include an .eps figure into your paper
%%
%\begin{figure}[htb]
%\centering
%\includegraphics[height=1.5in]{magnet}
%\caption{Plan of the magnet used in the mesmeric studies.}
%\label{fig:magnet}
%\end{figure}
%%%%%%%%%%%%%%%%%%%%%%%%%%%%%%%%%%%%%%%%%%%%%%%%%%%%%%%%%%%%%%%%%%%%%%%%%%%

\Acknowledgements

TB would like to thank the members of the LHCb collaboration. AI would like to thank the members of the Belle and Belle II collaborations. This work is supported in part by the Royal Society (TB),  the European Research Council under the advanced grant EFT4LHC and the cluster of excellence for 
precision physics, fundamental Interactions and structure of matter in Munich (DS). 

\clearpage

\bibliographystyle{LHCb}
\bibliography{references}

%\begin{thebibliography}{99}
%
%%%
%%%  bibliographic items can be constructed using the LaTeX format in SPIRES:
%%%    see    http://www.slac.stanford.edu/spires/hep/latex.html
%%%  SPIRES will also supply the CITATION line information; please include it.
%%%
%
%%\bibitem{Mesmer}
%%F. A. Mesmer, Proc. Wien. Acad. Sci. {\bf 13}, 1564, 1593 (1762).
%%%CITATION = PWASA,13,1564;%%
%
%
%\end{thebibliography}

\end{document}